\documentclass[prl,secnumarabic,amssymb, nobibnotes,aps,twocolumn]{revtex4-1}
\usepackage{graphicx , epsfig,color}
\usepackage{amsmath}
\usepackage{latexsym}
\usepackage{amstext}
\usepackage{array}
\usepackage{multirow}
\usepackage{changebar}
\usepackage{subfigure}
\usepackage{stackrel}
\usepackage{ulem}

\newcommand{\ds}{\displaystyle}
\newcommand{\beq}{\begin{eqnarray}}
\newcommand{\eeq}{\end{eqnarray}}
\newcommand{\beqq}{\begin{eqnarray*}}
\newcommand{\eeqq}{\end{eqnarray*}}
\newcommand{\p}{\partial}

\newcommand{\eps}{\varepsilon}

\newcommand{\x}{\mbox{\boldmath$x$}}

\newcommand{\n}{\mbox{\boldmath$n$}}

\newcommand{\y}{\mbox{\boldmath$y$}}

\newcommand{\sm}{\setminus}
\font\bb=msbm10 at 12pt
\def\rR{\hbox{\bb R}}

\begin{document}

\title{Reconstructing the gradient source position from  steady-state fluxes to small receptors}
\author{U. Dobramysl$^{1}$, D. Holcman$^{2}$}
\affiliation{  $^1$ Gurdon Institute, University of Cambridge, Cambridge CB2 1QN, United Kingdom $^2$ Ecole Normale Sup\'erieure, France and Mathematical Institute, University of Oxford, OX2 6GG, United Kingdom.}

\begin{abstract}
Recovering the position of a source from the fluxes of diffusing particles through small receptors allows a biological cell to determine its relative position, spatial localization and guide it to a final target. However, how a source can be recovered from point fluxes remains unclear. Using the Narrow Escape Time approach for an open domain, we compute the diffusion fluxes of Brownian particles generated by a steady-state gradient from a single source through small holes distributed on a surface in two dimensions. {We find that the location of a source can be recovered when there are at least 3 receptors and the source is positioned no further than 10 cell radii away}, but this condition is not necessary in a narrow strip. The present approach provides a computational basis for the first step of direction sensing of a gradient at a single cell level.
\end{abstract}
\maketitle

Sensing a molecular gradient made of cue concentration is the first step to transform cell positional information into a genetic specialization and a differentiation signal \cite{Wolpert}. During axonal growth and guidance, the growth cone (the tip of a neuron) uses the concentration of morphogens \cite{Goodhill1,reviewREingruber} to decide whether or not to continue moving, stop, turn right or left. Bacteria and spermatozoa in particular are able to orient themselves in a chemotaxis gradient \cite{Kaupp1,Kaupp2,Kaupp3,Kaupp4,Heinrich3,Heinrich5,Goldstein1,Goldstein2,Powers}.  However, how a cell senses an external gradient concentration depends on its ability to estimate the fluxes of cues. These fluxes have been computed assuming that cues are fully or partially absorbed uniformly at the surface of a detecting ball \cite{BergPurcell}. These computations are used to estimate the sensitivity of the local concentration.  This, however, is insufficient to establish the orientation of the source of the gradient. Our aim is to clarify how a cell, which is only a few microns in size, can detect the direction of a source.\\
The first step of differentiating left from right certainly has to involve the spatial difference in the binding flux of external cues. Our model for cell direction detection uses a reflecting disk covered with small receptors. The receptors are perfect absorbers for Brownian particles (cues), emanating from a point source. Computing the fluxes of Brownian particles to small targets is part of the Narrow Escape Theory \cite{PNAS,PPR2013,holcmanschuss2015}, but this theory cannot be applied directly to open and unbounded domains, because the mean passage time of particles to any small target is infinite. To avoid this difficulty, we neglect the receptor binding time of diffusing molecules and consider that cell sensing is possible via direct measurement of the diffusion flux. However, we do not account for any further cellular transduction cascade that translates receptor local activation into an internal signal. {A receptor-local memory mechansim is necessary in order to prevent loss of directional information on the gradient due to homogenization of the downstream transduced signal (concentration of a second messenger or surface molecules) inside the cell.} Therefore, asymmetric fluxes at the receptor level should lead to an asymmetrical transduction inside the cell. Hence, we do not replace receptors by a homogenized boundary condition that would render measurements of spatial flux differences impossible.\\
In this letter, we first compute the fluxes of diffusing molecules to small targets ($n=2$ and $3$) located on the surface of a detecting disk. We evaluate the effect of different receptor arrangements and also study the influence of an infinitely long, confining narrow strip. Secondly, we estimate the maximum distance from a source at which a given concentration can be detected with a pre-defined accuracy. Finally, we study how the location of a source can be recovered from the difference of fluxes.  We demonstrate that the source position of a gradient can be reconstructed with three receptors, while sensing of the concentration level can be achieved with two only. \\
{\noindent \bf Diffusion fluxes through narrow windows.} The probability density function $p_t(\x,\x_0)$ for a Brownian particle generated at location $\x_0$ that can be absorbed on the boundary of a detecting two-dimensional disk of radius $R$, $\Omega=D(R)$, by windows $\p\Omega_{1}\cup\hdots\cup \p\Omega_{N}$ located on the surface $\p \Omega$ of the disk $\Omega$ satisfies
{\small
\beq
\frac{\p p_t}{\p t}&=&D\Delta p_t  \label{eqDP2}  \\
p_t(\x,\x_0)& = & \delta(\x-\x_0) \;\text{ for }\; \x \in \rR^2\!-\!\Omega\text{ and }t=0 \nonumber\\
\nonumber \ds \frac{\p p_t}{\p \n}(\x,\x_0) & = & 0 \;\;\text{for}\;\; \x\, \in\, \p\Omega-(\p \Omega_{1}\cup\hdots \cup\p\Omega_{N})\\
 \nonumber p_t(\x,\x_0) & = & 0 \;\;\text{for}\;\; \x\, \in\, \p\Omega_{1}\cup\hdots\cup \p\Omega_{N}.
 \eeq
 }
The reflecting boundary condition at $ \p\Omega - (\p\Omega_{1}\cup\hdots\cup \p\Omega_{N})$ accounts for the impenetrable boundary so that diffusing molecules are reflected on the surface. The absorbing boundary condition on each of the windows $\p \Omega_{1}\cup\hdots \cup\p\Omega_{N}$ represents rapid binding with a diffusion limited activation rate. The window sizes are identical and equal to $|\p \Omega_{1}|=\eps\ll 1$. The steady-state probability density $P_0(\x)$ is computed by solving the mixed boundary value problem for the Laplace equation \cite{holcmanschuss2015}
\beq
-D\Delta P_0(\x) & = & \delta(\x-\x_0) \;\;\text{ for }\;\; \x\, \in \,\rR^2- \Omega \label{eqDP2b}\\
 \nonumber \ds \frac{\p P_0}{\p \n}(\x) & = & 0 \;\;\text{for}\;\; \x\, \in\, \p\Omega \sm (\p \Omega_{1}\cup\hdots \cup\p\Omega_{N})\\
 \nonumber P_0(\x) & = & 0 \;\;\text{for}\;\; \x\, \in\, \p\Omega_{1}\cup\hdots\cup \p\Omega_{N}.
 \eeq
Although the density $P_0(\x)$ {is non-normalizable in two dimensions},
we are only interested in the splitting probability between windows, i.e. the normalized steady-state flux at window $k$,
{\small \beq
P_k= \ds \frac{\ds \int_{\p\Omega_{k}} \ds \frac{\p P_0(\x)}{\p \n} dS_{\x}}{\ds \sum_{q} \int_{\p\Omega_{q}} \ds \frac{\p P_0(\x)}{\p \n} dS_{\x}}.
\eeq }
Due to the recurrent property of the Brownian motion in two dimensions, the probability to hit a window before going to infinity is one, thus
\beq
\sum_{q} \int_{\p\Omega_{q}} \ds \frac{\p P_0(\x)}{\p \n} dS_{\x}=1. \label{eqNormWins}
\eeq
The fluxes for $N=2$ windows can be computed using the Green's function $G(\x,\y)$ of the domain using matched asymptotic expansion~\cite{Coombs,Ward} and involves a Green's function Matrix in general. However, using identity Eq.~(\ref{eqNormWins}), it is sufficient to compute only one probability and we get
\beq \label{eq:disk}
P_2=\frac12+ \frac{\pi}{2}\frac{G(\x_1,\x_0)-G(\x_2,\x_0)}{\{\log|\x_1-\x_2|-\log \eps\}},
\eeq
where the external Neumann-Green's function of a disk $\Omega=D(R)$ of radius $R$, for $\x, \y \in \rR^2 -B(R)$ is
\beq \label{green}
G(\x,\y)=\frac{-1}{2\pi }\left(\ln |\x-\y| + \ln \left|\frac{R^2}{|\x|^2}\x-\y\right| \right).
\eeq\\
\begin{figure}[http!]
  \centering
  \includegraphics{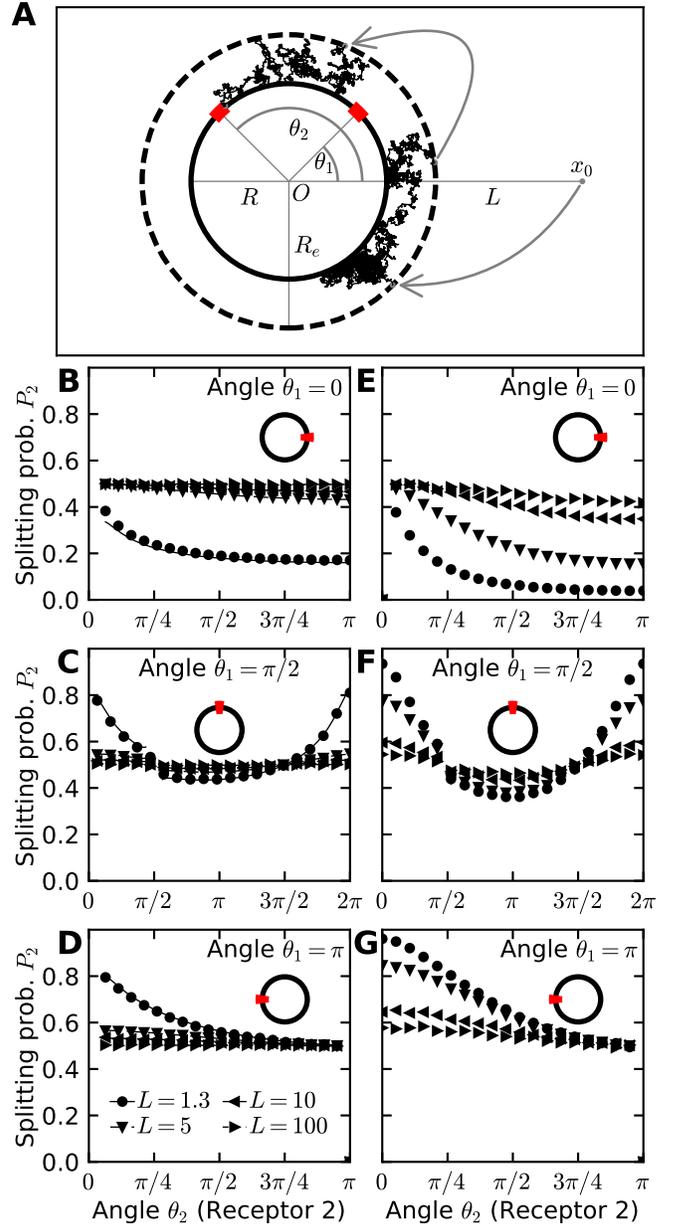}
  \caption{\textbf{Diffusion fluxes to small windows on the disk surface.} (A) Schematic representation of a mixed stochastic simulation of Brownian particles released at position $x_0$ at a distance $L=|x_0|$ from the origin $O$. Two windows of size $2\epsilon$ are placed on the circumference of the disk of radius $R$ in two dimensions or the equator of a sphere in three dimensions at angles $\theta_1$ and $\theta_2$ with the $x$-axis. Brownian particles are injected at a distance $R_{e}$ (dashed circle), as shown by the grey arrows. (B) Splitting probability (normalized flux) at window 2 {in two dimensions} with angle $\theta_1=0$, (C) $\theta_1=\pi/2$ (the jump in the analytical solution at $\pi/2$ emerges due to divergence when windows overlap), and (D) $\theta_1=\pi$. Simulations (markers) are compared to analytical solutions (solid lines). {(E) Splitting probability at window 2 in three dimensions (flux normalized to the total flux absorbed by any of the two windows) with angle $\theta_1=0$, (F) $\theta_1=\pi/2$, and (G) $\theta_1=\pi$.}}\label{fig:disk}
\end{figure}
To evaluate how the probability $P_2$ changes with the distance of the source $\x_0$ and the relative position of the windows, we compare Brownian simulations with the analytical expression~(\ref{eq:disk}) for a disk (Fig. \ref{fig:disk}A). For the simulations, we generated Brownian trajectories near the disk (on the surface of a disk of radius $R_e$) according to the exit point distribution of a process from an internal disk~\cite{schuss2013}. Interestingly, already at a distance of $L=10R$, the absolute difference between the fluxes $\Delta P=|P_1-P_2|$ is within $5\%$, making it almost impossible to determine source direction or concentration differences in a noisy environment.
The results are independent of the window positions and is qualitatively the same in dim 2 and 3 (compared Fig. \ref{fig:disk}B-C-D to E-F-G). Moreover,  $\Delta P\to 0$ as $L$ increases, see Fig. \ref{fig:disk}B-D. \\
{In contrast, when the disk is located in a narrow strip (Fig. \ref{fig:strip}A), the difference of fluxes between the two windows converges asymptotically to a finite difference $\Delta P$ depending on the strip width $a$, even for large source distances $L\geq 100R$ (see Fig. \ref{fig:strip}B-D.). Indeed, the fluxes hardly show any dependence on source distance $L$. The narrow funnel~\cite{PPR2013} between the strip and the disk prevents Brownian particles to reach a window located on the opposite side of the disk, leading to the observed effects.}\\
\begin{figure}[t!]
  \centering
  \includegraphics[scale=0.7]{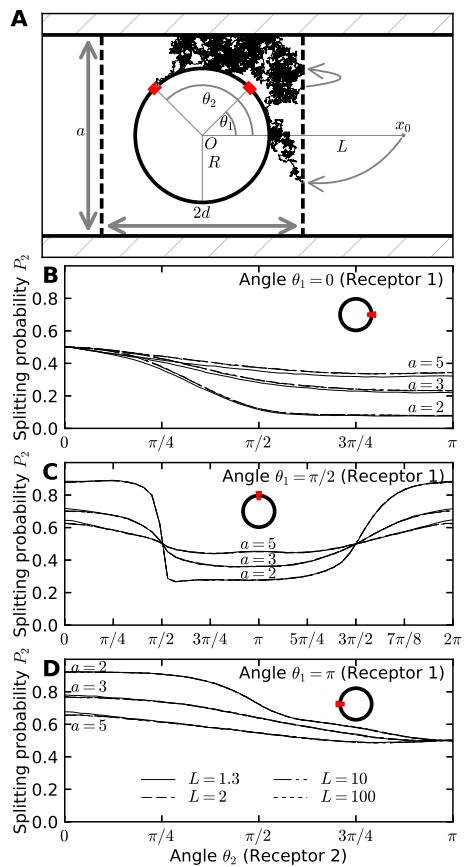}
\caption{\textbf{Diffusion fluxes to small windows for a disk in a narrow strip of width $a$.} (A) Scheme of the mixed stochastic simulations of Brownian particles confined in the strip and released at position $x_0$ at a distance $L=|x_0|$ from the origin $O$. Two windows of size $2\epsilon$ are placed on the circumference of the disk of radius $R$ at angles $\theta_1$ and $\theta_2$ with the $x$-axis. Brownian particles are injected at a distance $d$ on both sides of the disk (dashed vertical lines) and reflected from the strip walls at $y=\pm a/2$. (B) Splitting probability (normalized flux) at window 2 with angle $\theta_1=0$, (C) $\theta_1=\pi/2$, and (D) $\theta_1=\pi$.}
  \label{fig:strip}
\end{figure}
To further investigate how the window positions could influence the recovery of the source location, we estimated the maximum distance between the source and the disk containing two absorbing windows located at position $\x_1,\x_2$ that gives a significant difference of probability flux. For that purpose, we define the sensitivity ratio as
\beq \label{ratio1}
r(\x_1,\x_2,\x_0)=\frac{|P_1(\x_1,\x_2,\x_0)-P_2(\x_1,\x_2,\x_0)|}
{P_1(\x_1,\x_2,\x_0)+P_2(\x_1,\x_2,\x_0)}
\eeq
(note that here $P_1(\x_1,\x_2,\x_0)+P_2(\x_1,\x_2,\x_0)=1$). The domain of sensitivity for a threshold $T_{h}$ is the interior of the two-dimensional region
\beq\label{dsregion}
D_{S}=\{\x_0 \hbox{ such that } r(\x_1,\x_2,\x_0)\geq T_{h} \}.
\eeq
We plotted the boundary of the region $D_S$ for two absorbing windows symmetrically positioned (Fig.~\ref{fig:sens}A) and when the angle is $\theta_{12}=\pi/2$ (Fig.~\ref{fig:sens}B) for $T_{h}=1\%, 5\%$ and $10\%$. The region $D_{S}$ consists of two connected components and no detection (for a threshold below $T_h$) is possible outside $D_{S}$. Interestingly, with a $1\%$ precision, the domain is around 20$\times$ the size of the detecting disk. Beyond this distance, no detection is possible.\\
\begin{figure}[t!]
  \centering
 \includegraphics[scale=0.8]{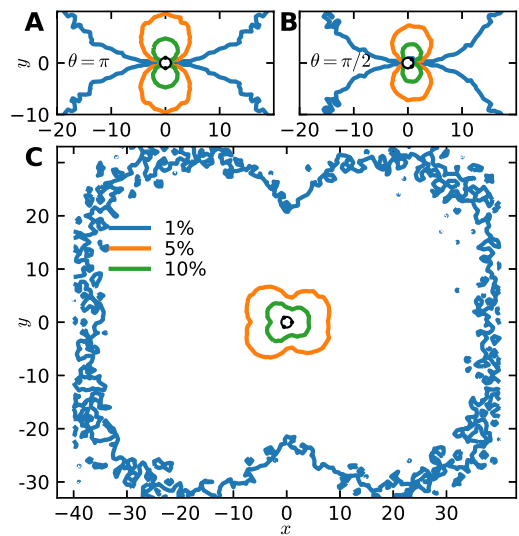}
  \caption{\textbf{Detectable region and contours for small windows on a disk}  (A) Two windows are placed on a disk with angular spacing $\theta=\pi$. The contours indicate the position for a threshold  $T_h=1\%$, $5\%$ and $10\%$, given by the normalized flux difference or probability~(\ref{ratio1}). (B) Two windows are placed with an angle $\pi/2$ apart. (C) Three windows placed $2\pi/3$ apart. The contours of $D(S)$ are given by $r_3=T_h$ (relation~(\ref{ratio2})). 
  } \label{fig:sens}
\end{figure}
{The detection sensitivity for a given source location $\x_0$ and optimal window placement is defined by}
\beq \label{diff}
f(\x_0)=\max_{\x_1,\x_2} |P_1(\x_1,\x_2,\x_0)-P_2(\x_1,\x_2,\x_0)|.
\eeq
The maximum is achieved for a window configuration aligned with the position of the source and symmetric with respect to the center of the disk
centered at the origin. An explicit computation with $\x_2=-\x_1$, $|\x_1|=|\x_2|=R$ gives
\beq \label{decay}
f(\x_0)= \ds\frac{2R}{|\x_0|\log\frac{2R}{\eps}}+o\left(\frac{1}{|\x_0|}\right),
\eeq
where $R$ is the radius of the disk. {Hence, the detection sensitivity decreases as the reciprocal of the distance to the source $L=|\x_0|$.} With three windows, detection is possible if at least one of the difference between the splitting probability is higher than the threshold $T_h$. {We thus define the new sensitivity using
\beq \label{ratio2}
r_3(\x_1,\x_2,\x_3,\x_0)=\frac{\max\{|P_1-P_2|,|P_1-P_3|\}}{P_1+P_2+P_3},
\eeq with $P_1$, $P_2$ and $P_3$ defined above depend on $\x_1$, $\x_2$, $\x_3$
and $\x_0$.} {Note that this definition illustrates the range of sensitivity and it might be possible that cells perform using biochemical reactions this computation.} The detectability region is now completely surrounding the detecting disk $\Omega$ and the boundaries are much larger than in the
two-windows case. Numerical simulations show the region $D_{S}$ defined in
Eq.~(\ref{dsregion}) for the function $r_3$ with windows positioned at the
corners of an equilateral triangle is now connected and seems to extend to 40
times the size of the detecting disk (Fig. \ref{fig:sens}C). Finally,
reconstructing the location of the source in the detectable region $D_S$ from
the steady-state probability fluxes, requires inverting Eq.~(\ref{eq:disk}) and
to find $\x_0$ when the fluxes are known. With two windows located on a
detecting disk and using the expression of the Green's function~(\ref{green}),
we obtain a one dimensional curve (Fig. \ref{fig:pos}A). {At least three windows are required to recover the point source, located at the intersection of two curves (Fig. 4B). Indeed, the recurrent Brownian motion in dimension 2 \cite{schuss2013} implies that $P_1+P_2+P_3=1$, thus we only need to estimate P1 and P2 from Eq.~\eqref{eqDP2b}, by the matched asymptotic method, except that now we have to invert a three-dimensional matrix equation (see SI). We use the expression Eq.~\eqref{green} to invert the equations and the intersection point is found numerically via MINPACK's multidimensional nonlinear root finding method
\textit{hybrj}~\cite{hybrj}. Interestingly, fluctuations in the probability fluxes $P_i$ (implemented by changing $P_i\to [1\pm\eta]P_i$) yields a nonlinear and spatially inhomogeneous uncertainty in the reconstruction of the source position (overlap of the shaded areas in Fig.~\ref{fig:pos}B).\\
\begin{figure}[t!]
  \centering
 \includegraphics[scale=0.7]{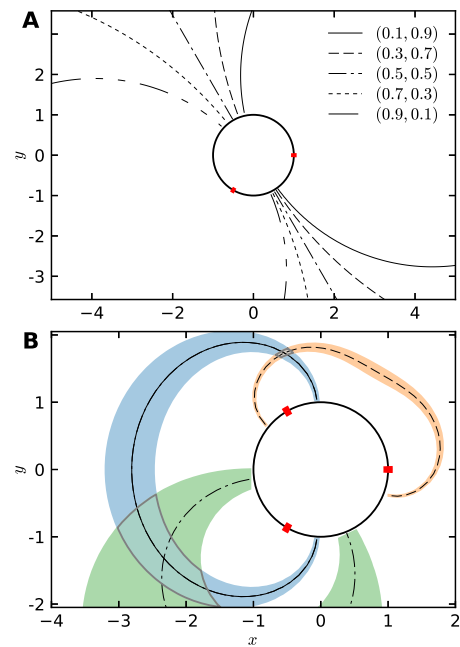}
  \caption{\textbf{Recovery a source position from fluxes at small windows}. (A) Two windows are positioned on a detecting disk with an angular spacing of $2\pi/3$. This arrangement allows recovery of the source position $\x_0$ located on a one dimensional curve. The curves are the ensemble of solutions for $\x_0$ computed from equation~(\ref{eq:disk}) for the fluxes $(P_1, P_2)$ displayed in the figure legend. (B) Intersection of two curves described computed from extending equation~(\ref{eq:disk}) to three windows from which the source position $\x_0$ is uniquely recovered. The dashed line was calculated for $P_1=0.13$, $P_2=0.05$ and $P_3=0.82$, while the dash-dotted line shows the case $P_1=0.13$, $P_2=0.58$ and $P_3=0.29$. The solid line does not change between the two cases. Shaded areas indicate the amplitude of the fluctuations for a fixed uncertainty $\eta=0.15$.} \label{fig:pos}
\end{figure}
{\noindent \bf Discussion and Conclusion.}  The ability of a cell to sense a gradient concentration is mediated by the binding of Brownian molecules to receptors~\cite{BergPurcell,HS2005}. {Computing the diffusion fluxes via homogenizing over local receptor positions~\cite{Endres,Goodhill1,Berez}, renders a recovery of any directional information impossible, as it assigns the same flux to the entire boundary of the detecting disk.} Based on Narrow Escape Theory~\cite{Ward,Coombs,holcmanschuss2015}, we estimated the probability fluxes on each individual receptor window separately and found that in two dimensions (i.e. a flat environment), the direction of the source can be recovered. {This mechanism requires a comparison between the fluxes of at least three boundary receptors.} In addition, the Green's function approach allows us to estimate the boundary of the region of sensitivity {characterized by a difference of fluxes between receptors being larger than a predefined sensitivity threshold.} The sensitivity decays with the reciprocal of the distance to the source according to Eq.~(\ref{decay}). Furthermore, we evaluated the effect of the external geometry on the threshold of sensitivity: although a disk of radius R with absorbing windows cannot sense the source position located {beyond ten cell radii}, in a narrow band, the detection is possible due to narrow passages for the diffusing molecules \cite{PPR2013}, {as long as one detecting window is facing away from the source.} \\
The present method can be used for several applications, such as growth-cone navigation inside the developing brain. Neurons have to travel millimeters to centimeters to find the correct cortical regions and to form synaptic connections~\cite{reviewREingruber,Goodhill1,Yogev}. We propose that { narrow extracellular tubes formed} of neurons and glial cells probably allows the genesis of shallow gradients detected by receptors located on the growth cone, although it is not known precisely how these receptors are organized. The dynamics of the growth cone including moving protrusions is certainly an additional mechanism worthy of further investigation in context of direction sensing, but should dominate for short-range distances only.\\
Although we focussed on two and three windows only, the results would be similar for two or three receptor clusters~\cite{holcmanschuss2015}. The model we used here was developed in the fast binding limit without rebinding \cite{Kaupp1}. It remains a challenge to apply the present theory to understand how bacteria \cite{Heinrich3}, sperm \cite{Kaupp2}  or neurite growth \cite{Yogev} localize the source of their cues. We limited the present approach to the initial level of source detection, however the asymmetry of receptor detection needs to persist in the downstream transduction, which certainly is another key question to investigate.\\


\end{document}